\documentclass[prd, preprint, aps, nofootinbib]{revtex4}

\usepackage{enumerate}	  
\usepackage{amsmath}
\usepackage{amsfonts}

\numberwithin{equation}{section}	
\newcommand{\be}{\begin{equation}}	\newcommand{\bea}{\begin{eqnarray}}
\newcommand{\eea}{\end{eqnarray}} \newcommand{\ee}{\end{equation}}
\newcommand{\non}{\nonumber}

			\def\l{\ell}				
\def\v{{\mathfrak v}}	\def\a{{\mathfrak a}}	 			 	
		\def\b{{\mathfrak b}}	\def\x{{\mathsf X}}	\def\M{{\cal M}}

\begin{document}

\begin{titlepage}

\vfill
\begin{center}
\baselineskip=16pt
{\Large\bf Painlev\'{e}-Gullstrand-type coordinates for the five-dimensional Myers-Perry black hole}

\vskip 0.3cm
{\large {\sl }}
\vskip 10.mm
{\bf ~Tehani K. Finch$^{\dagger}$ } \\

\vskip 1cm

{\small
NASA Goddard Space Flight Center\\
Greenbelt MD 20771\\
}
\end{center}
\vfill
\par

\begin{center}
{\bf ABSTRACT}
\end{center}
\begin{quote}

The Painlev\'{e}-Gullstrand coordinates provide a convenient framework for presenting the Schwarzschild geometry because of their flat constant-time hypersurfaces, and the fact that they are free of coordinate singularities outside r=0.  Generalizations of Painlev\'{e}-Gullstrand coordinates suitable for the Kerr geometry have been presented by Doran and Nat\'{a}rio. These coordinate systems feature a time coordinate identical to the proper time of zero-angular-momentum observers that are dropped from infinity. Here, the methods of Doran and Nat\'{a}rio are extended to the five-dimensional rotating black hole found by Myers and Perry. The result is a new formulation of the Myers-Perry metric.  The properties and physical significance of these new coordinates are discussed. 

\vfill	 \hrule width 5.cm   \vskip 2.mm
$^\dagger$ tehani.k.finch (at) nasa.gov

\end{quote}
\end{titlepage}

%%%%%%%%%%%%%%%%%%%%%%%%%%%%%%%%%%%%%%%%
\section*{Introduction}
\label{sec:intro}
By using the Birkhoff theorem, the Schwarzschild geometry has been shown to be the unique vacuum spherically symmetric solution of the four-dimensional Einstein equations.  The Kerr geometry, on the other hand, has been shown only to be the unique stationary, rotating vacuum \emph{black hole} solution of the four-dimensional Einstein equations.  No distribution of \emph{matter} is currently known to produce a Kerr exterior.  Thus the Kerr geometry does not necessarily correspond to the spacetime outside a rotating star or planet \cite{Vis08}.  This is an indication of the complications encountered upon trying to extend results found for the Schwarzschild spacetime to the Kerr spacetime. 

One such extension of particular relevance for the present investigation involves the Painlev\'{e}-Gullstrand form of the metric of the Schwarzschild black hole \cite{Pain21,Gull22}.
This metric has attracted renewed interest of late, due to its useful properties, which include the following: (i) its hypersurfaces of constant time are flat; (ii) its coordinates are well-behaved at the horizon; (iii) its time coordinate (which is timelike everywhere) coincides with the proper time of a freely-falling observer dropped from infinity.   
With these desirable characteristics, it would be natural to broaden this scheme to include the Kerr black hole, and approaches for doing so have been presented by Doran \cite{Dor00} and more recently by Nat\'{a}rio \cite{Nat08}.  These coordinate systems retain the feature (ii) and partially retain (iii) although, unsurprisingly, they do not possess trait (i). 

Given the current interest in black holes in dimensions higher than four, it is also of interest to consider the higher-dimensional analogs of the Doran and Nat\'{a}rio coordinate systems.  General Relativity in dimensions higher than four has added complexity; the black hole uniqueness theorems of four dimensions do not apply and event horizons with nonspherical topology can be realized.  A review of these black objects is given in, $e.g.$ \cite{Emp08}, and we shall reserve the term ``black hole" for objects with horizons of spherical topology.  The general higher-dimensional solutions of the Einstein equations that represent rotating black holes in a vacuum spacetime were found in 1986 by Myers and Perry \cite{MP86}.  
Our attention here will be focused on the specific case of a rotating black hole in 4+1 dimensions, which we term the Myers-Perry geometry.    

Painlev\'{e}-Gullstrand coordinates are presented in Section \ref{sec:PG}, and the relevant formulations of the Kerr geometry are reviewed in Section \ref{sec:Kerr}.  The extension to the Myers-Perry black hole begins in Section \ref{sec:5D}, and is given in Kerr-Schild coordinates in Section \ref{sec:KS}.  Section \ref{sec:equal} gives some results for the special case of Myers-Perry black holes with equal angular momenta.  In this paper, $c=1$ and a similar convention applies to the four-dimensional and five-dimensional gravitational constants: $G_4=1$ and $G_5=1$, except where otherwise specified.  Four-dimensional quantities will be taken to have Latin indices $(i,j,..)$ while five-dimensional quantities will have Greek indices $(\kappa,\nu,...)$. 
\section{Painlev\'{e}-Gullstrand coordinates}
\label{sec:PG}
The Schwarzschild solution is most often given as 
\be ds^2 = -(1-2M/r)dt_s^2 +\frac{dr^2}{1-2M/r} + r^2 d\theta^2 + r^2\sin^2\theta d\phi'^2.\ee
The conversion to Painlev\'{e}-Gullstrand (PG) coordinates involves the relation 
\be d\bar{t} = dt_s + \frac{\sqrt{2M/r}}{1-2M/r}dr, \ee 
so that the metric takes the form
\be \label{PG-4D} ds^2 = -d\bar{t}^2 +(dr-\bar{\v} d\bar{t})^2 + r^2 d\theta^2 + r^2\sin^2\theta d\phi'^2.  \ee
Here $\bar{\v}=-\sqrt{2M/r}$ is identical to the proper velocity of a particle dropped from rest at infinity in this geometry.\footnote{Strictly speaking, of course, the observer begins a large finite distance from the black hole.  The phrase ``at infinity" refers to the fact that certain physical quantities of such an observer have mathematically well defined limits as the observer's initial position becomes infinitely far from the black hole.}  
The proper time of such a particle corresponds to $\bar{t}$, so that the four-velocity of the particle is given by
\be \label{PGfourvelocity}	\frac{dx^i}{d\tau}:=\dot{x}^i = (1,\bar{\v},0,0). \ee
The time $\bar{t}$ is related to Schwarzschild time by $t_s$ by 
\be \bar{t} = t_s + 2\sqrt{2Mr} + 2M\ln \left| \frac{\sqrt{r/(2M)}-1}{\sqrt{r/(2M)}+1} \right| + C, \ee 
with $C$ being a constant of integration.  Setting $C=0$ implies that $\bar{t}=t_s$ at $r=0$. 
It can be seen from (\ref{PG-4D}) and the form of $\v$ that hypersurfaces of constant $\bar{t}$ are spatially flat and that there are no coordinate singularities away from $r=0$.  Thus these coordinates are well-behaved at the horizon, and indeed can be extended inward all the way to the singularity.
\section{Review of the Kerr solution}
\label{sec:Kerr}
One familiar form of the Kerr solution is that of Boyer-Lindquist (BL) coordinates $(t',r,\theta,\phi')$ \cite{Boyer67}:
\be ds^2=-(1-\frac{2M r}{\rho^2})dt'^2 + \frac{\rho^2}{\Delta}dr^2 + \rho^2 d\theta^2 +\Sigma\sin^2\theta d\phi'^2-\frac{4M r a \sin^2\theta}{\rho^2}dt' d\phi'  \ee
where 
\be \rho^2 = r^2 + a^2\cos^2\theta, \quad \Delta= r^2-2Mr+a^2 \quad\mbox{and} \quad \Sigma= r^2 + a^2+\frac{2M r a^2}{\rho^2}\sin^2\theta. \non \ee
It will be useful to rewrite this metric in the form 
\be ds^2 = -dt'^2 + \frac{ \rho^2}{\Delta}dr^2 + \rho^2 d\theta^2 + \frac{2M r}{\rho^2}(dt'-a\sin^2\theta d\phi' )^2 + (r^2+a^2)\sin^2\theta d{\phi'}^2. \ee
The outer stationary limit surface, or ergosurface, is given by outer solution of $g_{t't'}=0 \leftrightarrow \rho^2 = 2Mr$.  There are two radii at which $\Delta=0$; the outer one, $r_+$, is the event horizon and the inner one, $r_-$, is a Cauchy horizon. 
The curvature scalar $R_{ijkl}R^{ijkl}$ diverges at the ``ring singularity" specified by $\rho^2=0$, \emph{i.e.} $(r=0,\theta=\pi/2$).
The principal null directions of the Kerr metric in BL coordinates are given by
\be \label{pndbl}	\l_{\pm}^i\partial_i = \frac{r^2+a^2}{\Delta}\partial_{t'} \pm \partial_r+\frac{a}{\Delta}\partial_{\phi'} \, . \ee 
Photons for which $dx^i/ d\lambda = \l_{\pm}^i$ follow principal null geodesics, with the plus and minus signs corresponding to outgoing and ingoing directions, respectively.  

Due to the presence of the azimuthal killing vector in the Kerr geometry, any observer moving along a geodesic will have a conserved (azimuthal component of) angular momentum.  
Consider a freely falling observer (FREFO) whose trajectory starts from rest at infinity.  The conserved angular momentum of this observer is zero.   
In BL coordinates, such an observer has a proper velocity given by 
\be  \dot{x}^i = (1+ \frac{2Mr(r^2+a^2)}{\rho^2 \Delta},\v,0,\frac{2Mar}{\rho^2 \Delta}). \ee
A new form of the Kerr solution was found by Doran \cite{Dor00} in which the motion of such an observer takes the same form as (\ref{PGfourvelocity}).  In terms of the Doran coordinates $(t,r,\theta,\phi'')$ the metric is given by
\be \label{Dor-4D} ds^2 = -dt^2 +\frac{\rho^2}{r^2+a^2}[dr-\v(dt-a\sin^2\theta d\phi'')]^2 + \rho^2 d\theta^2 + (r^2+a^2)\sin^2\theta d\phi''^2  \ee
in which 
\be \v=-\frac{\sqrt{2Mr(r^2+a^2)}}{\rho^2} \ee
is the radial proper velocity of the FREFO.  The four-velocity of such a FREFO in Doran coordinates is 
\be \label{Dorfourvelocity}	\dot{x}^i = (1,\v,0,0). \ee
The utility of this coordinate system is seen in the physical significance of the coordinates $t$ and $\phi''$, with $t$ corresponding to the proper time of the FREFO dropped from infinity, and $\phi''$ constructed so that she has a radially directed four-velocity.  The principal null directions in Doran coordinates are given by
\be \label{pnddor}	\l_{\pm}^i\partial_i = (r^2+a^2)\partial_t  +\left( -\sqrt{2Mr(r^2+a^2)}\pm(r^2+a^2) \right)	\partial_r + a\partial_{\phi''} \, . \ee

%%%%%%%%%%
For future convenience, we now take the intermediate step of introducing the coordinates $(v,r,\theta,\Phi)$, in the spirit of the original work of Kerr \cite{Kerr63}.  This produces what is often called the advanced Eddington-Finkelstein form of the geometry: 
\bea ds^2 &=& -(1-\frac{2Mr}{\rho^2})(dv + a\sin^2\theta d\Phi)^2 +2(dv + a\sin^2\theta d\Phi)(dr + a\sin^2\theta d\Phi)\non \\ 
&+& \rho^2(d\theta^2 + \sin^2\theta d \Phi^2). \eea
After performing the substitution $v=T+r$ we obtain
\bea ds^2 &=& -dT^2+dr^2+2a\sin^2\theta dr d\Phi+\rho^2 d\theta^2+(r^2+a^2)\sin^2\theta d\Phi^2 \non \\
&+& \frac{2Mr}{\rho^2}(dT+dr+a\sin^2\theta d\Phi)^2.\label{kerrTr}\eea
These coordinates can be transformed into the BL coordinates via 
\be dT=dt'+\frac{2mr}{\Delta}dr, \hskip 2.0cm  d\Phi = -d\phi' - \frac{a}{\Delta}dr.		\ee
We now relate the Doran coordinates $(t,r,\theta,\phi'')$ to the advanced Eddington-Finkelstein coordinates  $(v,r,\theta,\Phi)$ as follows:
\bea dv &=&dt+\frac{(r^2+a^2) dr}{r^2+a^2-\rho^2 v } \\ 
d\Phi &=& -d\phi'' - \frac{a dr}{r^2+a^2 - \rho^2 v}. 
\eea
%%%%%%%%%%
Nat\'{a}rio introduced another form of the Kerr metric in \cite{Nat08} using the coordinates $(t,r,\theta,\phi)$:
\be \label{Nat-4D} ds^2 = -dt^2 + \frac{\rho^2}{\Sigma}(dr-\v dt)^2+\rho^2 d\theta^2+\Sigma\sin^2\theta(d\phi+\delta d\theta - \Omega dt)^2, \ee
for which
\be \Omega(r,\theta) = \frac{2Mar}{\rho^2\Sigma} 
\ee
is the familiar BL angular velocity $\frac{d\phi'}{dt'}$ 
of any observer with zero angular momentum at given $(r,\theta)$. 
The details of the derivation are omitted here, but can be found in the appendix of \cite{Nat08}.   
The function $\delta$, which vanishes for $r=+\infty$, is given by\footnote{Here we mention that the form of $\delta$ obtained here is the negative of that presented in \cite{Nat08}.} 
\be \delta(r,\theta) =-a^2\sin2\theta \int_r^{+\infty}\frac{\v\Omega}{\Sigma}d\tilde{r}, \ee
where $\tilde{r}$ has been introduced as a dummy variable of integration.  The coordinates are related through the transformations
\bea dt' &=& dt+\frac{\rho^2 \v}{\Delta}dr, \\ 
d\phi' &=& d\phi + \frac{\rho^2 \v}{\Delta}\Omega dr + \delta d\theta = d\phi''+\frac{a}{r^2+a^2}\frac{\rho^2 \v}{\Delta}dr. \eea
Both the Doran form of the Kerr metric in (\ref{Dor-4D}) and the Nat\'{a}rio form in (\ref{Nat-4D}) are well behaved at the horizons, when they exist.  In the limit of $a\rightarrow 0$ both the Doran and Nat\'{a}rio metrics reduce to the PG metric.  Namely, in such a limit, $\{t\rightarrow \bar{t}, \v\rightarrow\bar{\v}, \phi''\rightarrow\phi', \phi\rightarrow\phi' \}$. 

Interestingly, the use of the Painlev\'{e}-Gullstrand coordinates allows for an interpretation of a ``river model" \cite{HL08} in which one views space as flowing radially inward to the black hole through a flat background.  In the Schwarzschild geometry, the velocity of the river at a given point is the Newtonian escape velocity at that point.  The magnitude of the river velocity reaches the speed of light at the event horizon and exceeds the speed of light inside the horizon, making it impossible for objects or photons to stay still.  In the process of extending this scheme to the Kerr geometry, the Doran coordinates were used and a ``twist field" was introduced in \cite{HL08}.  The Nat\'{a}rio coordinate system provides an alternative this method.  In the Nat\'{a}rio framework there is no twist field, the overall magnitude of the river velocity reaches the speed of light at the ergosurface, and the \emph{radial} component of the river velocity reaches the speed of light at the event horizon.  

\section{Overview of the five-dimensional results}
\label{sec:5D}
The five-dimensional analog of the Schwarzschild solution is the D=5 version of the Tangherlini metric \cite{Tang63},
\be ds^2 = -(1-\M/r^2)dt_s^2 +\frac{dr^2}{1-\M/r^2} + r^2 d\theta^2 + r^2\sin^2\theta d\phi_1'^2+ r^2\cos^2\theta d\phi_2'^2,\ee  
where the parameter $\M$ is related to the mass $M$ via $\M=8G_5M/(3\pi)$ and $G_5$ is the five-dimensional gravitational constant.  The angles satisfy $0\le\theta\le \pi/2$ and $0\le\phi_1,\phi_2 < 2\pi $.  The metric can be written in PG-style coordinates thusly:
\be \label{PG-5D} ds^2 = -d\bar{t}^2 +(dr-\bar{\v}_5 d\bar{t})^2 + r^2 d\theta^2 + r^2\sin^2\theta d\phi_1'^2+ r^2\cos^2\theta d\phi_2'^2,  \ee
where $\bar{\v}_5=-\sqrt{\M/r^2}$ is again the proper velocity of a particle dropped from rest at infinity.  

Meanwhile, the four spatial directions give rise to two independent planes of rotation, and thus to two independent azimuthal coordinates, each with its associated angular momentum.  A rotating black hole in this spacetime is the five-dimensional version of the general solution discovered by Myers and Perry (MP)\cite{MP86}.  This  solution is characterized by a mass $M$ and two angular momenta $J_1$ and $J_2$.  It is convenient to introduce the parameters $a$ and $b$ such that \cite{Fro11} 
\be  	
a=\frac{3}{2}\frac{J_1}{M},  \hskip 2.0cm b=\frac{3}{2}\frac{J_2}{M}. \ee
When written in terms of BL-type coordinates $(t',r,\theta,\phi'_1,\phi'_2)$, the MP metric is as follows \cite{Fro03}: 
\bea ds^2 &=&-dt'^2+ \frac{r^2 \rho_5^2}{\Delta_5}dr^2 + \rho_5^2 d\theta^2 +\frac{\M}{\rho_5^2}(dt'-a\sin^2\theta d\phi'_1-b\cos^2\theta d\phi'_2)^2 \nonumber\\
&+& (r^2+a^2)\sin^2\theta d{\phi'_1}^2 + (r^2+b^2)\cos^2\theta d{\phi'_2}^2,
\eea
\be \mbox{where}\quad \rho_5^2 = r^2+a^2\cos^2\theta+b^2\sin^2\theta \quad\mbox{and}\quad \Delta_5=(r^2+a^2)(r^2+b^2)-\M r^2. \nonumber 
\ee

A shortcoming of these coordinates is that they do not they do not encompass the entire spacetime, and a coordinate singularity occurs at $r=0$.  It was pointed out in \cite{MP86} that if we label the angular momenta such that $a^2\ge b^2$, the introduction of the variable $\x=r^2$, where $\x \in [-b^2, \infty] $, remedies this problem.  The origin now corresponds to $(\x=-b^2,\theta=0)$.  The ring singularity, which is the genuine curvature singularity, occurs at $\rho_5^2=0$, $i.e.$ $(\x=-b^2,\theta=\pi/2)$.  As  with four dimensions, the ergosurface is given by $g_{t't'}=0$, $i.e.$ $\rho_5^2 = \M$.  The outer and inner horizons $r_\pm$ are the solutions of $\Delta_5=0$.   %
The principal null directions in BL coordinates are	
\be \label{pndmpbl}	n_{\pm}^{\nu}\partial_{\nu} = \frac{(r^2+a^2)(r^2+b^2)}{\Delta_5}\partial_{t'}
\pm \partial_r + \frac{a(r^2+b^2)}{\Delta_5}\partial_{\phi_1'}+\frac{b(r^2+a^2)}{\Delta_5}\partial_{\phi_2'}.
\ee
%%%%%%%%%%

We now construct new coordinates for the MP geometry, following a procedure analogous to that discussed for the Kerr case in Section \ref{sec:Kerr}.  Expressing the MP metric in advanced Eddington-Finkelstein form gives:  
\bea ds^2 &=& -(1-\frac{\M}{\rho_5^2})(dv + a\sin^2\theta d\Phi_1 + b\cos^2\theta d\Phi_2)^2 \\ &+&2(dv + a\sin^2\theta d\Phi_1 + b\cos^2\theta d\Phi_2)(dr + a\sin^2\theta d\Phi_1 + b\cos^2\theta d\Phi_2 )\non \\ 
&+& \rho_5^2 d\theta^2 + (r^2+a^2\cos^2\theta)\sin^2\theta d \Phi_1^2 + (r^2+b^2\sin^2\theta)\cos^2\theta d\Phi_2^2\\
&-&2ab\cos^2\theta\sin^2\theta d\Phi_1 d\Phi_2. \non 	\eea
The substitution $v=T+r$ yields
\bea ds^2 &=& -dT^2+dr^2+2dr (a\sin^2\theta d\Phi_1+ b\cos^2\theta d\Phi_2)\\
&+& \rho_5^2 d\theta^2+(r^2+a^2)\sin^2\theta d\Phi_1^2 + (r^2+b^2)\cos^2\theta d\Phi_2^2 \non \\
&+& \frac{\M}{\rho_5^2}(dT+dr+a\sin^2\theta d\Phi_1+ b\cos^2\theta d\Phi_2)^2. \non \label{mpTr}\eea
These coordinates can be transformed into the BL coordinates via 
\bea dT &=& dt'+\frac{\M r^2}{\Delta_5}dr, \\
d\Phi_1 &=& -d\phi_1' - \frac{a(b^2+r^2)}{\Delta_5}dr, \hskip 1.0cm d\Phi_2 = -d\phi_2' - \frac{b(a^2+r^2)}{\Delta_5}dr.	\non	\eea

%%%%%%%%%%
One of the main results of this investigation is the expression of the Myers-Perry metric in Doran-type coordinates:
\bea ds^2 &=& -dt^2+\rho_5^2\frac{r^2}{(r^2+a^2)(r^2+b^2)}(dr-\v_{5}(dt-a\sin^2\theta d\phi''_1-b\cos^2\theta d\phi''_2))^2 \nonumber \\
&+&  \rho_5^2 d\theta^2 + (r^2+a^2)\sin^2\theta d{\phi''_1}^2 + (r^2+b^2)\cos^2\theta d{\phi''_2}^2,
\label{Dor-5D} \eea
\be \mbox{where}\quad\quad\quad \v_5 = -\frac{\sqrt{\M(r^2+a^2)(r^2+b^2)} }{r\rho_5^2}. \nonumber \ee
In five dimensions,  
an freely falling observer dropped from from rest at infinity will have a proper velocity 
\be \frac{dx^{\nu}}{d\tau}:=\dot{x}^{\nu} = (1,\v_5,0,0,0). \ee
Since $dt/d\tau=1$ we see that these are a privileged class of observers whose proper time corresponds to coordinate time $t$.
The principal null directions in Doran coordinates are given by
\bea 	
n_{\pm}^{\nu}\partial_{\nu} &=& \frac{(r^2+a^2)(r^2+b^2)}{r^2}\partial_t+\left(-\frac{\sqrt{\M(r^2+a^2)(r^2+b^2)} }{r} \pm(r^2+a^2+b^2+\frac{a^2 b^2}{r^2}) \right)\partial_r \non \\
&+&a(1+\frac{b^2}{r^2})\partial_{\phi_1''} + b(1+\frac{a^2}{r^2})\partial_{\phi_2''}. \label{pndmpdor} 
\eea

%%%%%%%%%%%%%%
The transformation from $(v,r,\theta,\Phi_1,\Phi_2)$ to the Doran coordinates has the form
\bea dv &=& dt+\frac{(r^2+a^2)(r^2+b^2)dr}{(r^2+a^2)(r^2+b^2) - r^2\rho_5^2 \v_5} \\
d\Phi_1 &=& -d\phi_1'' - \frac{a(r^2+b^2)dr}{(r^2+a^2)(r^2+b^2) - r^2\rho_5^2 \v_5} \non \\
d\Phi_2 &=& -d\phi_2'' - \frac{b(r^2+a^2)dr}{(r^2+a^2)(r^2+b^2) - r^2\rho_5^2 \v_5}.\non
\eea
Meanwhile, the MP geometry in Nat\'{a}rio-type coordinates has the form 
\bea ds^2 & = & -dt^2 + \frac{\rho_5^2}{\Xi}(dr - \v_5 dt)^2 + \rho_5^2 d\theta^2 + \lambda_1(d\phi_1+\delta_5 d\theta - \Omega_1 dt)^2 \label{Nat-5D} \\
&+& \lambda_2(d\phi_2 + \zeta_5 d\theta -\Omega_2 dt)^2 + 2\varpi(d\phi_1 +\delta_5 d\theta - \Omega_1 dt) (d\phi_2 + \zeta_5 d\theta -\Omega_2 dt), \nonumber
\eea
in which we have
\bea \delta_5(r,\theta) &=& (b^2-a^2)\sin2\theta \int_r^{+\infty}\frac{\v_5\, \Omega_1}{\Xi}d\tilde{r} \quad\mbox{and}\quad \\
 \zeta_5(r,\theta) &=& (b^2-a^2)\sin2\theta \int_r^{+\infty}\frac{\v_5\, \Omega_2}{\Xi}d\tilde{r},  \eea
again using $\tilde{r}$ as a dummy variable.  We have introduced the following quantities:
\bea 
\lambda_1 &=& (r^2+a^2)\sin^2\theta + \frac{a^2 \M \sin^4\theta}{\rho_5^2}, \nonumber \\
\lambda_2 &=& (r^2+b^2)\cos^2\theta + \frac{b^2 \M \cos^4\theta}{\rho_5^2}, \nonumber \\
\varpi &=& \frac{a b \M \cos^2\theta \sin^2\theta}{\rho_5^2}, \nonumber \\
\varsigma &=& 
(r^2+a^2)(r^2+b^2)\rho_5^2+\M b^2(r^2+a^2)\cos^2\theta + \M a^2 (r^2+b^2)\sin^2\theta,  \nonumber \\
\Xi &=& \frac{\lambda_1\lambda_2-\varpi^2}{r^2\sin^2\theta\cos^2\theta} = \frac{\varsigma}{r^2 \rho_5^2}, \nonumber \\
\Omega_1 &=& \frac{a \M (r^2+b^2)}{\varsigma},\nonumber \\
\Omega_2 &=& \frac{b \M (r^2+a^2)}{\varsigma}, \nonumber
\eea 
with $\Omega_1$ and $\Omega_2$ representing the $\phi_1$ and $\phi_2$ angular velocities 
of a FREFO.  Both the Doran-like form (\ref{Dor-5D}) and Nat\'{a}rio-like form (\ref{Nat-5D}) of the metric are well behaved at the horizons, but both also require the substitution $\x=r^2$ with $\x \in [-b^2, \infty] $ to span the full spacetime.  

These three sets of coordinates share the same $r$ and $\theta$ but have differing combinations of time and azimuthal coordinates, namely $(t',\phi'_1,\phi'_2)$, $(t,\phi''_1,\phi''_2)$ and $(t,\phi_1,\phi_2)$.  These are related through the transformations
\bea dt' &=& dt+\frac{r^2 \rho_5^2 \v_5}{\Delta_5}dr, \\ 
d\phi_1' &=& d\phi_1 + \frac{r^2 \rho_5^2 \v_5}{\Delta_5}\Omega_1 dr + \delta_5 d\theta = d\phi_1''+\frac{a}{r^2+a^2} \frac{r^2 \rho_5^2 \v_5}{\Delta_5}dr, \\
d\phi_2' &=& d\phi_2 + \frac{r^2 \rho_5^2 \v_5}{\Delta_5}\Omega_2 dr + \zeta_5 d\theta = d\phi_2''+\frac{b}{r^2+b^2} \frac{r^2 \rho_5^2 \v_5}{\Delta_5}dr.\eea 
The stationarity and bi-azimuthal symmetry of the geometry are evident in each of these systems.  
\section{Kerr-Schild-like form of the metrics}
\label{sec:KS}
A metric written such that 
\be g_{ij} = \eta_{ij} + 2Hk_i k_j \ee
is in Kerr-Schild form \cite{Kerr65}.  The Kerr metric itself can be cast in such a form using several different coordinate systems; see $e.g.$ \cite{Vis08}.
One of the better-known Kerr-Schild decompositions uses Cartesian-type 
coordinates $(T,x,y,z)$.  These are related to the coordinates $(T,r,\theta,\Phi)$ of (\ref{kerrTr}) by the relations
\begin{align}
x & = (r\cos\Phi+a\sin\Phi)\sin\theta= \sqrt{r^2+a^2}\sin\theta\cos(\Phi-\arctan(a/r)), \\
y & = (r\sin\Phi-a\cos\Phi)\sin\theta= \sqrt{r^2+a^2}\sin\theta\sin(\Phi-\arctan(a/r)), \\
z & = r\cos\theta, 
\end{align}
so that
\be \frac{x^2+y^2}{r^2+a^2}  +\frac{z^2}{r^2} =1. \label{r-4D} \ee
The Kerr metric is obtained by setting 
\be H=\frac{Mr^3}{r^4+a^2 z^2},\hskip 1.0cm k_i = \left(1, \frac{rx-ay}{r^2+a^2}, \frac{ry+ax}{r^2+a^2}, \frac{z}{r} \right).\ee
Doran's version of the Kerr metric has a representation in Cartesian-type coordinates somewhat reminiscent of the Kerr-Schild form.  The coordinates $(t,x'',y'',z'')$ are related to Doran's $(t,r,\theta,\phi'')$ using
\begin{align}
x'' & = \sqrt{r^2+a^2}\sin\theta\cos\phi'', \\
y'' & =\sqrt{r^2+a^2}\sin\theta\sin\phi'', \\
z'' & = r\cos\theta. 
\end{align}
The Kerr metric can be written in these coordinates as \cite{Dor00}
\be \label{Dor-KS}
g_{ij} = \eta_{ij} + \frac{2 \sqrt{2Mr}}{\rho^2} \a_{(i} V_{j)} +\frac{2Mr}{\rho^2} V_i V_j ,
\ee
where $\eta_{ij}$ is the 4D Minkowski metric.
\footnote{Reference \cite{Dor00} used (+,-,-,-) signature, whereas (-,+,+,+) is used here.} 
In the above we have
\begin{align}
V_i &= \left( 1, \frac{ay''}{r^2+a^2}, \frac{-ax''}{r^2+a^2},0 \right), 
\label{Vi} \\
\a_i &= (r^2+a^2)^{1/2} \left( 0, \frac{rx''}{r^2+a^2}, \frac{ry''}{r^2+a^2}, \frac{z''}{r} \right), \label{ai}
\end{align}
with $r$ defined implicitly via (\ref{r-4D}), and our symmetrization convention is such that
 \be \a_{(i}\b_{j)}: = \frac{1}{2}(\a_i \b_j + \a_j \b_i).\ee  
The principal null directions of this geometry $\l_{\pm}$ can be related to $V_i$ and $\a_i$ by
\begin{equation} 
\l_{i\,\pm} = (r^2+a^2)^{1/2} V_i \pm (\sqrt{2Mr} V_i + \a_i).
\label{pnd}
\end{equation} 

It turns out that analogous results hold for 
the Myers-Perry solution.  We first introduce the Cartesian-type coordinates $(T,x,y,z,w)$ such that
\begin{align}
x & = (r\cos\Phi_1+a\sin\Phi_1)\sin\theta= \sqrt{r^2+a^2}\sin\theta\cos(\Phi_1-\arctan(a/r)), \\
y & = (r\sin\Phi_1-a\cos\Phi_1)\sin\theta= \sqrt{r^2+a^2}\sin\theta\sin(\Phi_1-\arctan(a/r)), \\
z & = (r\cos\Phi_2+b\sin\Phi_2)\cos\theta= \sqrt{r^2+b^2}\cos\theta\cos(\Phi_2-\arctan(b/r)), \\
w & = (r\sin\Phi_1-b\cos\Phi_2)\cos\theta= \sqrt{r^2+b^2}\cos\theta\sin(\Phi_2-\arctan(b/r)),
\end{align}		
which have the property that
\be \frac{x^2+y^2}{r^2+a^2}  +\frac{z^2+w^2}{r^2+b^2} =1. \label{r-5D}	\ee
The metric expressed in Kerr-Schild form,
\be g_{\kappa\nu} = \eta_{\kappa\nu} + 2Hk_{\kappa} k_{\nu}, \ee
is such that
\bea H &=& \frac{\M r^2}{r^4+a^2(w^2+z^2)+b^2(x^2+y^2)-a^2 b^2},\\ 
k_{\nu} &=& \left(1, \frac{rx+ay}{r^2+a^2}, \frac{ry-ax}{r^2+a^2}, \frac{rz+bw}{r^2+b^2}, \frac{rw-bz}{r^2+b^2} \right).\eea  
Moving to five dimensions, the Doran form of the MP metric can be written in a form analogous to that in (\ref{Dor-KS}).  After introducing 
\begin{align}
x'' & = \sqrt{r^2+a^2}\sin\theta\cos\phi_1'', \\
y'' & = \sqrt{r^2+a^2}\sin\theta\sin\phi_1'', \\
z'' & = \sqrt{r^2+b^2}\cos\theta\cos\phi_2'',\\
w'' & = \sqrt{r^2+b^2}\cos\theta\sin\phi_2'',
\end{align}
the result is
\be
g_{\kappa\nu} = \eta_{\kappa\nu} + \frac{2 \sqrt{\M}}{\rho^2_5} \b_{(\kappa}W_{\nu)} +
\frac{\M}{\rho_5^2} W_{\kappa}W_{\nu}\, ,
\label{KS-MP}
\ee
in which $\eta_{\kappa\nu}$ is the 5D Minkowski metric, and the parameters $\b_{\nu}$ and $W_{\nu}$ satisfy
\begin{align}
W_{\nu} &= \left(1, \frac{a y''}{r^2+a^2}, \frac{-ax''}{r^2+a^2}, \frac{b w''}{r^2+b^2}, \frac{-b z''}{r^2+b^2} \right) \quad\mbox{and} \label{V5} \\
\b_{\nu} &= \frac{\sqrt{(r^2+a^2)(r^2+b^2)}}{r} \left( 0, \frac{r x''}{r^2+a^2}, \frac{r y''}{r^2+a^2},\frac{r z''}{r^2+b^2}, \frac{r w''}{r^2+b^2} \right). \label{amu}
\end{align}
Meanwhile, the principal null directions are given by
\be 
n_{\nu\pm} = \frac{\sqrt{(r^2+a^2)(r^2+b^2)}}{r}W_{\nu}  \pm (\sqrt{\M}W_{\nu}  + \b_{\nu}).
\label{pnd-mp}
\end{equation} 
As discussed above for BL coordinates, if $a$ and $b$ are nonzero, these $(x,y,z,w)$ do not extend inside the spheroid given by $\frac{x^2+y^2}{a^2}+\frac{z^2+w^2}{b^2}=1$, $i.e.$ inside the region $r=0$; an analogous statement holds for $(x'',y'',z'',w'')$.
\section{The case of equal angular momenta}
\label{sec:equal}
Five-dimensional MP black holes with $a^2=b^2\neq 0$, $i.e.$ $J_1=J_2\neq 0$, have more symmetry than the generic ones; their horizons take the form of ``squashed" three-spheres. 
Setting $b^2=a^2$ enhances the spatial symmetry from $U(1)\times U(1)$ to $U(2)$. 
There is no analog of a squashed three-sphere in lower dimensions, so this situation affords us the opportunity to consider new phenomena.\\
We begin with the form of the metric, taking $b=+a=\frac{3J}{2M}$.  In BL-type coordinates we obtain
\bea ds^2 &=&-dt'^2+ \frac{r^2 (r^2+a^2)}{\tilde{\Delta}_5}dr^2 +\frac{\M}{(r^2+a^2)}(dt'-a(\sin^2\theta d\phi'_1+\cos^2\theta d\phi'_2))^2 \nonumber\\
&+& (r^2+a^2)(d\theta^2+\sin^2\theta d{\phi'_1}^2 + \cos^2\theta d{\phi'_2}^2),
\eea
where $\tilde{\Delta}_5=(r^2+a^2)^2-\M r^2$.
In Doran-type coordinates the metric becomes
\bea ds^2 &=& -dt^2+\frac{r^2}{r^2+a^2}\Big(dr-\tilde{\v}_{5}(dt-a(\sin^2\theta d\phi''_1+\cos^2\theta d\phi''_2)\,)\Big)^2 \nonumber \\
&+&  (r^2+a^2) (d\theta^2 +\sin^2\theta d{\phi''_1}^2 + \cos^2\theta d{\phi''_2}^2),
\eea
and we see that $\v_5$ has reduced to $\tilde{\v}_{5}=-\sqrt{\M}/r$, which has no dependence on $\theta$. \\
Perhaps the simplification is most striking in Nat\'{a}rio-type coordinates, because there, when $b=a$, $\delta_5$ and $\zeta_5$ both vanish:
\bea ds^2 & = & -dt^2 + \frac{(r^2+a^2)}{\tilde{\Xi}}(dr - \tilde{\v}_5 dt)^2 + (r^2+a^2) d\theta^2 + \tilde{\lambda}_1(d\phi_1 - \Omega dt)^2 \nonumber \\
&+& \tilde{\lambda}_2(d\phi_2 - \tilde{\Omega} dt)^2 + 2\tilde{\varpi}(d\phi_1  - \tilde{\Omega} dt) (d\phi_2  - \tilde{\Omega} dt),
\eea
where we now have
\bea 
\tilde{\lambda}_1 &=& (r^2+a^2)\sin^2\theta + \frac{a^2 \M \sin^4\theta}{(r^2+a^2)}, \nonumber \\
\tilde{\lambda}_2 &=& (r^2+a^2)\cos^2\theta + \frac{a^2 \M \cos^4\theta}{(r^2+a^2)}, \nonumber \\
\tilde{\varpi} &=& \frac{a^2 \M \cos^2\theta \sin^2\theta}{(r^2+a^2)}, \nonumber \\
\tilde{\varsigma} &=& (r^2+a^2)( (r^2+a^2)^2 + \M a^2),  \nonumber \\
\tilde{\Xi} &=& \frac{\tilde{\lambda}_1\tilde{\lambda}_2-\tilde{\varpi}^2}{r^2\sin^2\theta\cos^2\theta}  
=\frac{(r^2+a^2)^2 + \M a^2}{r^2}, \nonumber \\
\tilde{\Omega} &=& \frac{a \M (r^2 +a^2)}{\varsigma} = \frac{a \M}{(r^2+a^2)^2 + \M a^2}.\nonumber 
\eea 
Like $\tilde{\v}_5$, $\tilde{\Omega}$ has no $\theta$ dependence.\\
The relations between the coordinate systems $(t',\phi'_1,\phi'_2)$, $(t,\phi''_1,\phi''_2)$ and $(t,\phi_1,\phi_2)$ also simplify when $b=a$:
\bea dt' &=& dt-\frac{r(r^2+a^2)\sqrt{\M}}{\tilde{\Delta}_5}\,dr, \\ 
d\phi_1' &=& d\phi_1 -\frac{r(r^2+a^2)\sqrt{\M}}{\tilde{\Delta}_5}\tilde{\Omega} dr  = d\phi_1''- \frac{a r\sqrt{\M}}{\tilde{\Delta}_5}\,dr, \\
d\phi_2' &=& d\phi_2 -\frac{r(r^2+a^2)\sqrt{\M}}{\tilde{\Delta}_5}\tilde{\Omega} dr  = d\phi_2''- \frac{a r\sqrt{\M}}{\tilde{\Delta}_5}dr.\eea 
Here too, the $\theta$-dependence has vanished.  
%
%%%%%%%%%%%%%%%%%%%%%%%%%%%%%%
\section*{Conclusion}
\label{sec:conclusion}
Doran in \cite{Dor00} and Nat\'{a}rio in \cite{Nat08} have found ways of extending some of the virtues of the Painlev\'{e}-Gullstrand coordinate system to the Kerr geometry.  It has been demonstrated herein that it is possible to extend each system to the 4+1 Myers-Perry geometry, although the presence of two independent angular momenta makes this extension nontrivial.  As expected, the slices are not spatially flat but they do penetrate the horizon, and the time coordinate corresponds to the proper time of an object dropped from infinity.  The extension of these coordinate systems to spacetimes with charge $Q$ and/or cosmological constant $\Lambda$ is not the subject of the present work, but we note that Lin and Soo \cite{LS13} have argued that such an extension would require the use of an additional adjustable function $f(r)$ in the coordinate transformation.  In recent years Doran coordinates have found use in describing phenomena in the vicinity of rotating black holes in 3+1 dimensions that include Hawking radiation \cite{LS13}, $e^-e^+$ pair creation \cite{Ch12}, and EPR correlations \cite{Said10}.  Thus, the fact that analogous coordinate systems can be constructed in 4+1 dimensions offers promise for alternative and potentially useful viewpoints on higher-dimensional black hole physics.
  
\begin{acknowledgments} 
The author gratefully acknowledges support from the Department of Physics and Astronomy at Howard University; support from a NASA Postdoctoral Fellowship from the Oak Ridge Associated Universities; and discussions with James Lindesay.  
\end{acknowledgments} 

\appendix
\section{Tetrad and pentad} 
In \cite{Dor00} several tetrads and inverse tetrads with respect to the Doran coordinates $(t,r,\theta,\phi'')$ were introduced. Among these was the following inverse tetrad $(\tilde{\omega}^0, \tilde{\omega}^1, \tilde{\omega}^2, \tilde{\omega}^3  )$: 
\begin{align}
{\omega^0}_i dx^i &= dt, \\ 
{\omega^1}_i dx^i &=  \frac{\sqrt{2Mr}}{\rho}dt+ \frac{\rho}{\sqrt{r^2+a^2}}dr -\frac{\sqrt{2Mr}}{\rho} a \sin^2\!\theta \, d\phi''	,\\
{\omega^2}_i dx^i &= \rho \,d\theta ,  \\
{\omega^3}_i dx^i &= \sqrt{r^2+a^2} \sin\!\theta \, d\phi'' \, . 
\end{align}
Such a basis 
satisfies the relation ${\omega^K}_i {\omega^L}_j \eta_{KL} = g_{ij}$, where $\eta_{KL}$ = $diag(-1,1,1,1)$.  
The corresponding tetrad $(\vec{e}_0,\vec{e}_1, \vec{e}_2, \vec{e}_3)$ is 
\begin{align}
{e_0}^i \partial_i &= \partial_t -\frac{ \sqrt{2Mr(r^2+a^2)} }{\rho^2}\partial_r\, , \\
{e_1}^i \partial_i &= \frac{\sqrt{r^2+a^2}}{\rho}\partial_r\, ,  \\
{e_2}^i \partial_i &= \frac{1}{\rho}\partial_\theta\, ,   \\
{e_3}^i \partial_i &= \frac{\sqrt{2Mr}a\sin\theta }{\rho^2}\partial_r + \frac{1}{\sin\theta\sqrt{r^2+a^2}}\partial_{\phi''}\, ,  
 \end{align}
and the vectors satisfy $\vec{e}_K \cdot \vec{e}_L = \eta_{KL}$.

This inverse tetrad has an analogous Myers-Perry inverse pentad $(\tilde{\omega}^0, \tilde{\omega}^1, \tilde{\omega}^2, \tilde{\omega}^3, \tilde{\omega}^4 )$: 
\begin{align}
{\omega^0}_{\nu}dx^{\nu} &= dt,  \\ 
{\omega^1}_{\nu}dx^{\nu} &= \frac{\sqrt{\M}}{\rho_5}\, dt + \frac{r\rho_5}{\sqrt{(r^2+a^2)(r^2+a^2)} }dr -\frac{\sqrt{\M}}{{\rho_5}} a \sin^2\!\theta \,d\phi_1''  -\frac{\sqrt{\M}}{\rho_5} b \cos^2\!\theta \, d\phi_2'' ,		\\
{\omega^2}_{\nu}dx^{\nu} &=  \rho_5 \, d\theta,  \\
{\omega^3}_{\nu}dx^{\nu} &=  \sqrt{r^2+a^2} \sin\theta\, d\phi_1'' \, ,	 \\
{\omega^4}_{\nu}dx^{\nu} &=  \sqrt{r^2+b^2} \cos \theta\, d\phi_2'' \, , 
\end{align}
whose corresponding tetrad $(\vec{e}_0,\vec{e}_1, \vec{e}_2, \vec{e}_3, \vec{e}_4)$ is given by 
\begin{align}
{e_0}^{\nu} \partial_{\nu} &=\partial_t -\frac{ \sqrt{\M(r^2+a^2)(r^2+b^2)} }{r\rho_5^2}\partial_r\, , \\
{e_1}^{\nu} \partial_{\nu} &= \frac{\sqrt{(r^2+a^2)(r^2+b^2)}}{r\rho_5}\partial_r\, ,   \\
{e_2}^{\nu} \partial_{\nu} &= \frac{1}{\rho_5}\partial_\theta,   \\
{e_3}^{\nu} \partial_{\nu} &= \frac{a\sqrt{\M(r^2+b^2)}\sin\theta }{r\rho_5^2}\partial_r + \frac{1}{\sin\theta\sqrt{r^2+a^2}}\partial_{\phi_1''}\, ,   \\
{e_4}^{\nu} \partial_{\nu} &= \frac{b\sqrt{\M(r^2+a^2)}\cos\theta }{r\rho_5^2}\partial_r + \frac{1}{\cos\theta\sqrt{r^2+b^2}}\partial_{\phi_2''}.  
 \end{align}

\end{document}